\documentstyle[twocolumn,epsfig,prb,aps,amsmath,amssymb]{revtex}
\begin{document}

\twocolumn[\hsize\textwidth\columnwidth\hsize\csname
@twocolumnfalse\endcsname

\title{Charge and spin ordering, and charge transport properties
in a two-dimensional inhomogeneous $t-J$ model}

\author{Jos\'e A. Riera}
\address{
Instituto de F\'{\i}sica Rosario, Consejo Nacional de 
Investigaciones 
Cient\'{\i}ficas y T\'ecnicas, y Departamento de F\'{\i}sica,\\
Universidad Nacional de Rosario, Avenida Pellegrini 250, 
2000-Rosario, Argentina}
\date{\today}
\maketitle
\begin{abstract}
We study a two-dimensional $t-J$ model close to the Ising
limit in which charge inhomogeneity is stabilized by an on-site
potential $e_s$, by using diagonalization in a restricted Hilbert
space and finite temperature Quantum Monte Carlo. Both site and bond
centered stripes are considered and their similitudes and differences
are analyzed. The amplitude of charge
inhomogeneity is studied as $e_s\rightarrow 0$. Moreover, we show
that the anti-phase domain ordering occurs at a much lower
temperature than the formation of charge inhomogeneities and charge
localization. Hole-hole correlations indicate a metallic behavior
of the stripes with no signs of hole attraction. Kinetic energies
and current susceptibilities are computed and indications of charge
localization are discussed. The study of the doping dependence in the
range $0.083 \le x\le 0.167$ suggests that these features are
characteristic of the whole underdoped region.
\end{abstract}


\vskip2pc]
\section{Introduction}

After many years of intensive research the physics of underdoped
high-T$_c$ cuprates is still far from being wholly understood.
This underdoped region, where strongly correlated electron
effects are most important, are considered by many to contain
the key to explain central features of these materials,
including the superconducting phase. One of the main features
of the underdoped cuprates is the presence of the pseudogap
which has been intensely studied with several experimental
techniques.\cite{timusk,tallon} In spite of this intense effort,
there is still not a clear picture of the origin of the pseudogap
and its relation with the superconducting gap.\cite{PGnotS}
Another interesting phenomena observed by experiments in some
underdoped cuprates is the presence of ordered charge inhomogeneities
(``stripes") followed (as the temperature is decreased) by the
appearance of incommensurate (IC) magnetic
correlations.\cite{tranquada,ichikawa,mook} These charge
inhomogeneities provide a natural explanation for the pseudogap.
There is also strong experimental evidence of the presence of
charge ordering/stripes in other transition metal oxides like
manganites and nickelates.\cite{manganite,nickelate}
Again, in spite of an enormous amount of research, there is not
yet a clear picture of the mechanism leading to stripe formation
and the relation between stripes and superconductivity in the
cuprates. Different and in many cases competing scenarios have
been proposed to explain these
issues.\cite{emery,white,zaanen,castroneto}

From the theoretical point of view, in addition to the conceptual
complexity of strongly correlated electron systems, there are
important methodological difficulties. In stripes theories,
phenomenological models of coupled Luttinger (or Luther-Emery)
liquids have been studied\cite{dimcross} providing useful insight in
the problem although quantitative estimations are usually lacking.
In any case, these phenomenological models have to be supported
by the study of models at a more microscopic level. These
microscopic models (variants of Hubbard, $t-J$ models) are 
most frequently studied by numerical methods which also have to
deal with severe limitations (some of them will be discussed in
Section~\ref{model}). It is then absolutely necessary to complement
different approaches and to contrast the results obtained by
analytical and numerical techniques in order to get a consistent
picture of the physics of underdoped cuprates.

In this article a model of stripes is studied by numerical techniques.
The model is defined as a two-dimensional (2D) $t-J$ model where
stripes are stabilized by an effective on-site potential representing
the effect of extrinsic features to Cu-O planes like lattice
anisotropy,\cite{spatial} effect of chains in 
YBa$_2$Cu$_3$O$_{7-\delta}$ (Ref.~\onlinecite{mook}) or other
structural details,  Coulomb interaction due to off-plane charges, 
electron-phonon coupling,\cite{castroneto} etc. This model has been
studied numerically on small clusters ($4\times 4$) at both
zero\cite{eroles,prelovsek} and finite temperatures.\cite{shibata}.
Results on larger clusters with periodic boundary conditions were
obtained in Ref.~\onlinecite{riera}, although at the price of
working close to the Ising limit of the Heisenberg term of the
Hamiltonian. In the present article we present an exhaustive
analysis of this model by using quantum Monte Carlo (QMC) techniques,
in particular, we examine the behavior at various doping and the
effects of the anisotropy of the exchange term of the Hamiltonian.
These analysis at finite temperature are complemented with zero
temperature studies with a diagonalization technique in a reduced
Hilbert space known as ``systematically expanded Hilbert space"
(SEHS).\cite{sehs} The results obtained with both QMC and SEHS allow
a comparison not only with experiment but also with previous
calculations on similar clusters but with (partially) open boundary
conditions.\cite{white,martins}

The paper is organized as follows. In Section~\ref{model}, the model
considered is defined and some elementary details of the techniques
employed are discussed. In Section~\ref{stripfil} we present results
obtained with QMC and SEHS for the stripe filling as a function
of the on-site potential both for site- and bond-centered stripes
(defined in  Section~\ref{model}).
In Section~\ref{Torder} we study charge and spin ordering as
a function of temperature and in Section~\ref{invkin} we show the
behavior of the kinetic energy and current susceptibility with 
temperature as an indication
of charge transport properties in the underdoped region. Finally,
in the Conclusions we summarize the results as an unified picture
of the evolution of the various physical quantities examined as
a function of temperature.

\section{Model and methods.}
\label{model}

The Hamiltonian of the $t-J$ model with an anisotropic exchange term is:
\begin{eqnarray}
H_{tJ} =
&-& t \sum_{ \langle { i j} \rangle,\sigma }
({\tilde c}^{\dagger}_{ i\sigma}
{\tilde c}_{ j\sigma} + h.c. ) 
\nonumber  \\
&+& J \sum_{ \langle { i j} \rangle }
( \frac{\gamma}{2} (S^{+}_{i} S^{-}_{j} +
S^{-}_{i} S^{+}_{j}) +
S^{z}_{i} S^{z}_{j} -
{\frac{1}{4}} n_{i} n_{j} )
\label{ham_anis}
\end{eqnarray}
\noindent
where the notation is standard. $i,j$ are sites on a square lattice.
$N$ site clusters with fully periodic boundary conditions
(PBC) were considered.
We adopted $t=1$ as the unit of energy
and temperature and $J=0.35$. In some cases we considered $J=0.7$, 
specially because it reduces the sign problem. $\gamma=0$ corresponds 
to the Ising limit or $t-J_z$ model.\cite{tjz} The stripes are
stabilized by an effective on-site potential:
\begin{eqnarray}
H_{str}  = \sum_{i} {e_s}_i n_{i}
\label{stripepot}
\end{eqnarray}
\noindent
On ``site centered" (SC) stripes ${e_s}_i=-e_s < 0$ on equally
spaced vertical columns, as in the original picture of
Ref.~\onlinecite{tranquada}, and ${e_s}_i=e_s$ on the remainder
sites. At each doping fraction, the number of stripes imposed is such
that the linear hole density on the stripes at infinite potential is
1/2. We shall come to this point later on this section.

The QMC method employed is an implementation of the usual world-line
algorithm with the checkerboard decomposition\cite{reger,riera}.
The ``minus sign" problem\cite{minus} becomes more
severe as lattice size and doping are increased and temperature is
decreased. It is absent if holes are confined to move in one
dimension {\it and} the XY tern in the Heisenberg interaction is set
to zero. The behavior of the average sign as a function of various
parameters of the model is shown in Fig. 1 of
Ref.~\onlinecite{riera}. We checked our code against for the
$4\times 4$ cluster with one site-centered stripe, 2 holes,
$J/t=0.35$, $e_s=1$. We extrapolated our results to zero temperature
and compared them against the ones obtained with Lanczos exact
diagonalization. The relative error for the energy was of the order
of $10^{-4}$, and for correlations $10^{-3}$. For larger clusters,
smaller $e_s$, or larger $\gamma$ the correlation's error estimated
from
different independent runs may be as large as 10~$\%$ at the lowest
temperature attainable. In general, we kept $\Delta=\beta/M=0.083$,
where $\beta=1/T$ and $M$ is the Trotter number. For the lowest
temperatures of our simulations ($T/t \approx 0.1$) the measurement
runs were as long as 1.6~10$^6$ MC sweeps. With these very long
simulations we were able to deal with $\langle sgn \rangle$ as small
as 0.05. On the largest clusters we considered, $8\times 8$ and
$12\times 12$, we studied $e_s\approx 2$, which corresponds to 
well-defined charge inhomogeneities (see next section), and has
been also taken in previous related studies.\cite{eroles,prelovsek}

SEHS is a zero temperature diagonalization in a Hilbert space that is
expanded by successive hole hoppings, starting from an initial state
(e.g. holes on the stripes). After each expansion the Hamiltonian is
diagonalized with the Lanczos algorithm and a fraction of the most
weighted configurations of the ground state is kept for the next
iteration.\cite{sehs} The goal is that after several back and forth
steps, an optimal set of states is selected for a given dimension of
the Hilbert space. The Hilbert space is grown to a maximum
dimension (determined of course by the computational resources
available) and finally the quantities of interest are extrapolated
to the full dimension of the Hilbert space of the system. In
many cases reliable qualitative behaviors are obtained without
this final extrapolation. The whole procedure is variational but
it shares many of the virtues of exact diagonalization like the
possibility of computing any physical quantity both static and
dynamical. Another advantage that SEHS shares with Lanczos
diagonalization is the possibility of working for a given set of
quantum numbers, in particular since periodic boundary conditions
can be used, it is possible to work at a given momentum. In
addition there are no constraints on the model studied. In the
following section we will apply this technique to the $t-J_z$
model but in principle it could be applied to the isotropic
exchange although in this case requirements of time and memory
are much larger. Another possibility offered by this technique
is that long-range Coulomb interactions can be handled, as a
difference with QMC. We will explore in the future this possibility.

The use of the on-site potential to stabilize stripes could
be thought as leading to ``static" stripes, not in the sense of
``pinned" (since we are working with periodic boundary conditions)
but in the sense of suppressing stripe fluctuations. The opposite
meaning of ``static" could be applied to the stripes resulting from
density matrix renormalization group (DMRG) calculations\cite{white}
with open boundary conditions along the longer direction of the
cluster.  In this case, it is well known that ``bond-centered" (BC)
stripes are
obtained rather than SC ones. This fact may reflect the also
well known tendency of holes to form pairs in the 2D $t-J$ model
at zero temperature.\cite{rierayoung} However, SC stripes are much
sharper features than BC stripes and they seem more consistent with
elastic neutron scattering, NQR and ARPES\cite{zhou} data at low
doping although not definite conclusions can be drawn yet. At
higher doping there have been speculations that more BC stripes may
be formed at the expense of SC stripes thus explaining some
experimental features.\cite{zhou2}

\section{Stripe filling.}
\label{stripfil}

Our first study concerns the evolution of charge inhomogeneity as a
function of the applied on-site potential in the 2D $t-J_z$ model. 
Zero temperature results
were obtained with SEHS for a $8\times 4$ cluster, 4 holes 
(hole doping fraction $x=1/8$),
PBC. Stripes were imposed along the shorter direction ($y$ axis).
Translations along both directions, reflection along an horizontal
axis and spin reversal symmetries were imposed and the dimension
of the Hilbert space was grown up to $\approx 8.6~10^6$. The
ground state in the case of SC
stripes corresponds to $(q_x,q_y)=(0,\pi)$ and to $(0,0)$ in the
BC case.

\begin{figure}
\begin{center}
\epsfig{file=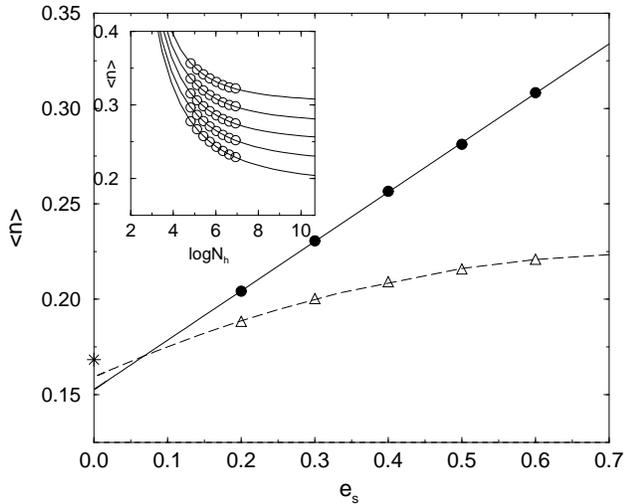,width=6.8cm,angle=-90}
\end{center}
\caption{Hole density on stripe sites on the $8\times 4$ cluster,
$J/t=0.35$, as a function of the on-site potential $e_s$. Circles
(triangles) correspond to SC (BC) stripes. Solid and dashed lines
are quadratic extrapolations to $e_s=0$. The star indicates the
result obtained with DMRG\cite{gazza} for (almost) BC stripes
on the same system but mixed boundary conditions. The inset shows
the stripe densities as a function of the dimension of the Hilbert
space and their extrapolation to the full dimension. From
bottom to top $e_s=0.2,\ldots,0.6$.
}
\label{fig1}
\end{figure}

Results for the stripe filling are depicted in
Fig.~\ref{fig1} for $J/t=0.35$ as a function of the on-site
potential. Details of the extrapolation to the full Hilbert space
dimension for each $e_s$ with a power law are shown in the inset of
this figure. These densities are then extrapolated with a
quadratic polynomial to $e_s=0$. The final extrapolated values of
the stripe densities are very similar (0.153 for SC and 0.159 for
BC stripes) and they are quite close to those obtained with DMRG in
Refs.~\onlinecite{white,martins}. Of course the {\em linear} density
in the BC case would correspond to the double of that quantity,
but this concept is somewhat misleading because it hides its relative
smoothness with respect to SC stripes. It has been also previously
shown that the effect of exchange anisotropy is not very
important.\cite{martins} We also include for comparison the DMRG
result for the same model, cluster and parameters but using mixed 
boundary conditions (open (periodic) along the direction perpendicular
(parallel) to the stripes).\cite{gazza} In this case, the occupancy of
the two legs on a stripe is slightly different and their average value
was taken. As expected, the open boundary conditions along the
direction transversal to the stripes leads to a somewhat
larger charge inhomogeneity than our results with fully PBC.

\begin{figure}
\begin{center}
\epsfig{file=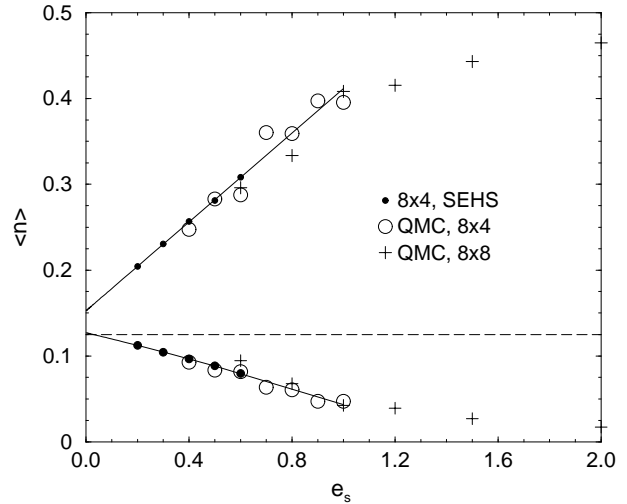,width=6.8cm,angle=-90}
\end{center}
\caption{Hole density on stripe sites on the $8\times 4$ and
$8\times 8$ clusters, $J/t=0.35$, as a function of the on-site
potential $e_s$ favoring a SC stripe. The dashed line corresponds
to the uniform doping $x=1/8$. Points above this line correspond
to the stripe; below it, they correspond to the first column next
to the stripe. Solid lines are quadratic fitting to the data.
}
\label{fig2}
\end{figure}

In the case of SC stripes we can compare the previous results with
the ones obtained with QMC. The temperature dependence of the stripe
density is very smooth\cite{riera} and it has been extrapolated to
zero temperature with a power law.
The stripe hole density on the $8\times 4$ cluster obtained with QMC
is shown in Fig.~\ref{fig2}. The resulting error of the finite
temperature data and of the extrapolation procedure can be appreciated
from the dispersion of the data. Nevertheless, the overall agreement
between QMC results and the ones obtained with SEHS is certainly very
good. The results for a larger cluster, $8\times 8$, are quite
consistent with the previous ones. The stripe density seems somewhat
smaller than for $8\times 4$, which could be explained by the
geometric anisotropy of this cluster making it more
``one-dimensional" thus favoring the formation of a charge density
wave along the long axis. However, error bars can likely mask this
effect, if it exists.

From the results of Fig.~\ref{fig2} one can also infer that to
have a stripe filling close to $1/2$, let us say, larger than 0.4,
the effective on-site potential should be  $\approx 1$ in the
case of SC stripes, and somewhat smaller ($\approx 0.35$) in the
case of BC stripes. Then, a sizable on-site potential, specially
in the case of SC stripes is necessary to stabilize this charge
inhomogeneity. A possible mechanism, not included explicitly in
our model, an in-plane long range Coulomb repulsion,\cite{emery}
could help to reduce the requirement of an external agent in order
to make stripes stable. A long-range potential cannot be implemented
in the world-line QMC used here (it cannot be implemented in DMRG
calculations either) but it could be included in the SEHS scheme as
we noticed in the previous section.

\section{Charge and spin ordering.}
\label{Torder}

There is now mounting experimental evidence that spin ordering,
signalled by incommensurate (IC) peaks in the magnetic structure
factor, follows charge ordering as temperature
decreases.\cite{tranquada,fujita,hunt} That is, charge degrees of 
freedom are the driving force of stripe formation. Similar 
behavior occurs in stripe formation in nickelates\cite{nickelate}
and manganites.\cite{manganite}
In the model here considered, in spite of a sizable charge
inhomogeneity at high temperatures, IC magnetic structure 
appears at much lower temperatures in agreement with experiment.
This is a nontrivial result, since some theoretical models of 
stripe formation do not properly account for this ordering
sequence.\cite{martins,zaanen}

\begin{figure}
\begin{center}
\epsfig{file=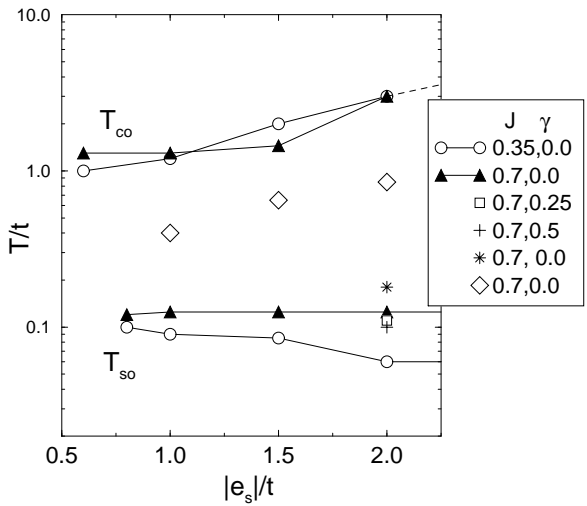,width=6.8cm,angle=-90}
\end{center}
\caption{ Phase diagram in the temperature-on-site potential plane
obtained for the $8\times 8$ cluster, $x=1/8$.
$T_{CO}$ and $T_{SO}$ correspond to the crossovers in the charge
and spin sectors discussed in the text. The point indicated with a
star was obtained on the $12\times 12$ cluster, $x=0.083$).
Diamonds indicate the temperature at which the inverse kinetic energy
along to the stripes are minimum.
}
\label{fig3}
\end{figure}

As an introduction to this section,
we reproduce the phase diagram in the temperature-e$_s$ plane
obtained in our previous work\cite{riera} for SC stripes
(Fig.~\ref{fig3}). By measuring charge and spin static structure
factors ($C({\bf k})$ and $S({\bf k})$, respectively),
two crossovers 
have been determined. At high temperature there is a crossover
in the charge sector determined by a change in the peak of
$C({\bf k})$ from $(0,0)$ to $(2 \delta,0)$ ($\delta=2 x\pi$).
At a much lower temperature, there is a
crossover in the spin sector signalled by the peak of $S({\bf k})$
splitting from $(\pi, \pi)$ to $(\pi\pm \delta, \pi)$, which
correspond to the peaks seen in neutron scattering experiments.
Between these two crossovers, typical temperature or energy
scales related to transport properties are located, as we will
discuss in Section~\ref{invkin}. These temperature scales have been
adopted to experimentally determine the onset of the
pseudogap.\cite{tallon}

The change of the peak in $C({\bf k})$ in our model is just a
consequence of stripe formation: $(2\delta,0)$ corresponds to the
main Fourier component of a pure SC stripe (away from $(0,0)$). The
change of the peak in $S({\bf k})$ corresponds to the
formation of anti-phase domains,\cite{tranquada} a feature which
is reproduced in our model.\cite{riera}
In the following subsections we will more
systematically study the effect of the anisotropy in the exchange
term of the Hamiltonian and the doping dependence of the results for
SC stripes. In the last subsection, we report similar studies
for BC stripes.

\subsection{Effect of the XY term.}

Although an antiferromagnetic (AF) correlation in the direction
perpendicular to the Cu-O planes would imply as a second order
process an effective exchange anisotropy in the plane, it is clear
that the most interesting model to study is the one that involves an
isotropic (or nearly isotropic) Heisenberg term. Unfortunately, a
more isotropic exchange implies stronger minus sign problem and 
hence, we are limited to consider $\gamma \le 0.5$. However,
as we show below,
one can get an insight on how and to what extent a more isotropic
coupling changes the results obtained on the Ising limit.

\begin{figure}
\begin{center}
\epsfig{file=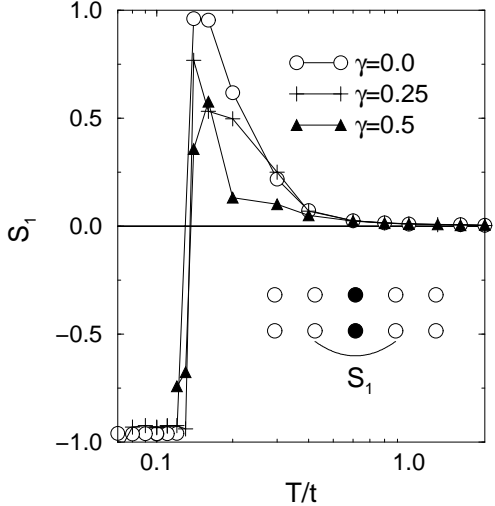,width=7.0cm,angle=-90}
\end{center}
\caption{Spin-spin correlation across the stripe (shown in the
inset) on the $8\times 8$
cluster, $x=1/8$, $J=0.7$, $e_s=1.5$, and several values of the
XY term $\gamma$.
}
\label{fig4}
\end{figure}

Let us consider the hole-hole and spin-spin correlations defined by
\begin{eqnarray}
C({\bf i,j}) = \langle n_{\bf i} n_{\bf j} \rangle
\end{eqnarray}
\noindent
and
\begin{eqnarray}
S({\bf i,j}) = \langle S^z_{\bf i} S^z_{\bf j} \rangle
\end{eqnarray}
respectively. $n_{\bf i}$ is the hole occupation number at site
${\bf i}$; ${\bf i}=(x,y)$ and ${\bf j}=(x^\prime,y^\prime)$. In
particular, if a SC stripe along $y$ is located at $x=x_s$, we are
interested in the correlations along the stripe, where
${\bf i}=(x_s,0)$, ${\bf j}=(x_s,r)$, and in correlations
across the stripe, corresponding to 
${\bf i}=(x_s-a,y)$, ${\bf j}=(x_s+a,y)$
(the inset of Fig.~\ref{fig4} shows the case of $a=1$).

In Fig.~\ref{fig4}, the spin-spin correlation across a vertical
stripe (indicated with filled circles in the inset) is shown for
the $8\times 8$ cluster, $x=1/8$,
$J=0.7$, $e_s=1.5$, and $\gamma=0$, 0.25, and 0.5. These correlations
show the change

\noindent
of sign corresponding of the spin domains going from
in-phase to anti-phase at the magnetic crossover temperature. Their
amplitude decrease as expected with increasing $\gamma$ but the
crossover temperature remains roughly unaffected. In this, and in
in similar figures, typical error bars are about the size of the
largest symbols used.

\begin{figure}
\begin{center}
\epsfig{file=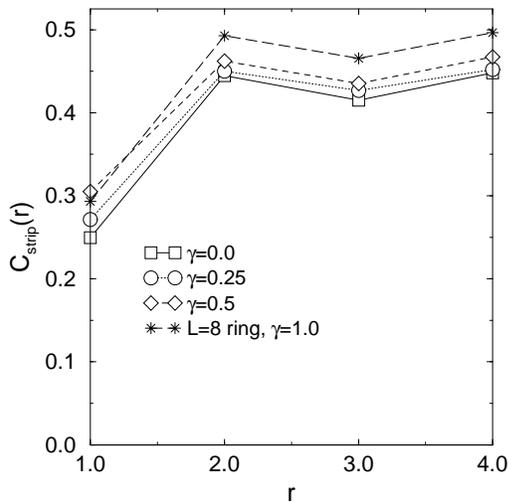,width=6.8cm,angle=-90}
\end{center}
\caption{Hole-hole correlations along the stripe on the $8\times 8$
cluster, $x=1/8$, $J=0.7$, $e_s=1.5$, and several values of the
XY term $\gamma$. Results obtained by exact diagonalization on
the quarter filled 8-site ring at zero temperature are also shown
for comparison.
}
\label{fig5}
\end{figure}

Another important result of Ref.~\onlinecite{riera}, viz. the metallic
behavior of the hole-hole correlations along a stripe is also
quite robust against the increase of the XY term of the exchange
interaction. In Fig.~\ref{fig5}, we show these correlations for
the $8\times 8$ cluster, $x=1/8$, $J=0.7$, $e_s=1.5$, and $\gamma=0$,
0.25, and 0.5 extrapolated to zero temperature. It can be seen that
their behavior remains unchanged
with $\gamma$ except for a uniform shift, within error bars (of the
order of the symbol size). This shift corresponds to a larger
filling of the stripe as $\gamma$ increases.

\subsection{Doping dependence.}

The study of the dependence of the results with doping is essential
to determine the extent to which the present mode is physically
realistic.
For this study, we performed QMC simulations on the $12\times 12$
cluster, with 12 holes ($x=0.083$) and 24 holes ($x=0.167$). In
the first case, two SC stripes six lattice spacings apart were
stabilized by an on-site potential, and in the second case four SC
stripes three lattice spacings apart. We have in addition results
for the $8\times 8$ cluster with eight holes and for the
$12\times 12$ cluster with 18 holes both corresponding to $x=0.125$.
In the latter cluster, care should be taken because the three
SC stripes present imply that the three intervening spin regions
cannot be in anti-phase.

\begin{figure}
\begin{center}
\epsfig{file=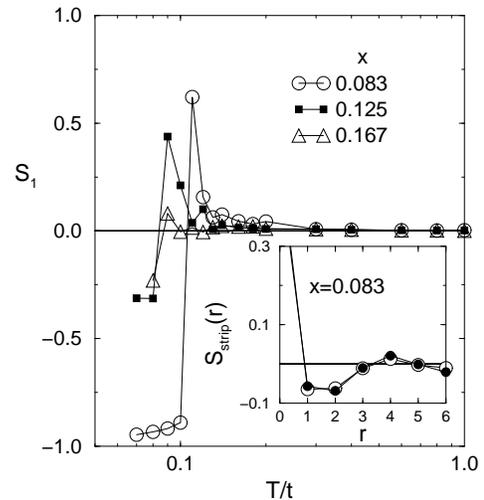,width=6.8cm,angle=-90}
\end{center}
\caption{Spin-spin correlations on the $12\times 12$ cluster,
$x=1/8$, $J=0.35$, $e_s=2.0$, $\gamma=0$, at several dopings
indicated on the text. The inset shows the spin-spin correlation
along the stripe vs. distance (open circles) at $T=0.09$ compared 
with exact results on a quarter filled 12 sites
chain (filled circles) at the same temperature.
}
\label{fig6}
\end{figure}

The spin-spin correlations across the stripe are shown in
Fig.~\ref{fig6}. The amplitude of the magnetic order both above and
below the crossover temperature decreases with doping and so it
does the crossover temperature itself, which is in agreement with
experiments. At $x=0.125$, for the $8\times 8$ cluster, the
crossover temperature is larger than that for the $12 \times 12$
one (due to the frustration mentioned above), but smaller than the
corresponding at $x=0.083$ (see
Fig.~3). The inset shows $S(r)$ along the stripe
for the same parameters, at the lowest reachable
temperature ($T=0.09$), and the corresponding results for a
chain obtained by exact diagonalization at the same temperature,
showing a remarkable agreement between them.

\begin{figure}
\begin{center}
\epsfig{file=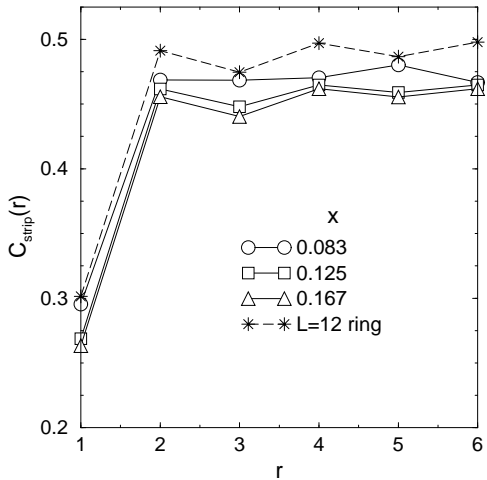,width=6.5cm,angle=-90}
\end{center}
\caption{Hole-hole correlations along a stripe on the $12\times 12$
cluster, $x=1/8$, $J=0.35$, $e_s=2.0$, $\gamma=0$. Results obtained
by exact diagonalization on the 12 site chain at 
zero temperature are also shown for comparison.
}
\label{fig7}
\end{figure}

The hole-hole correlations along a stripe, extrapolated to zero
temperature, are shown in Fig.~\ref{fig7} for the same parameters as
in Fig.~\ref{fig6}. These correlations also present the $4k_F$
oscillation of a  quarter-filled chain seen in Fig.~\ref{fig4}
with no signs of hole attraction. This behavior, together with 
the above mentioned similarity of spin-spin correlations indicate
that SC stripes behave at low temperatures like an isolated
chain.

\subsection{Bond-centered stripes.}

Bond centered stripes are stabilized by applying ${e_s}_i=-e_s < 0$
on equally spaced stripes consisting of two nearest neighbor columns
and ${e_s}_i= e_s > 0$ otherwise.
In this case, the sign problem, which is much worst than for
SC stripes, imposes us further limitations on the accessible
parameters and temperatures. Hence, this study was done for
the $8\times 8$ cluster at $x=1/8$ filling and on the Ising
limit of the Hamiltonian.

If the stripe legs along the $y$ direction are located at $x=x_s$
and $x_s+1$, then the correlations across the stripe involve
the sites ${\bf i}=(x_s-1,y)$, ${\bf j}=(x_s+2,y)$. The rung
correlations relate the sites ${\bf i}=(x_s,y)$,
${\bf j}=(x_s+1,y)$. The correlations along the stripes may
correspond to the same leg (${\bf i}=(x_s,0)$,
${\bf j}=(x_s,r)$) or to opposite legs (${\bf i}=(x_s,0)$,
${\bf j}=(x_s+1,r)$).

\begin{figure}
\begin{center}
\epsfig{file=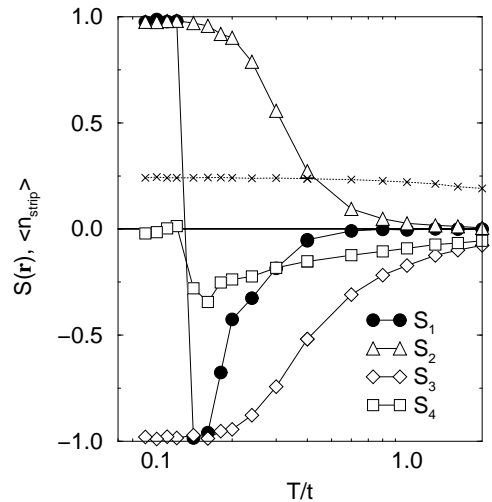,width=6.8cm,angle=-90}
\end{center}
\caption{Spin-spin correlations on the $8\times 8$ cluster,
$x=1/8$, $J=0.7$, $e_s=1.5$, $\gamma=0$, with bond centered stripes. 
$S_i, i=1,\ldots,4$ are defined in the text. The hole density on the
stripe, $<n_{str}>$ is shown with crosses.
}
\label{fig8}
\end{figure}

Our first important result is related to the appearance of
anti-phase spin ordering of the intervening domains at a temperature
at which the charges are already essentially moving on the stripes,
in complete analogy to what happens for SC stripes and in contrast
with some speculations based on an spin-only model.\cite{tworzydlo}
This result can be inferred from the spin-spin correlations shown
in Fig.~\ref{fig8}, obtained for $J=0.7$, $e_s=1.5$, $\gamma=0$.
For $J=0.35$, $e_s \le 2$, the sign problem inhibits us to reach
low enough temperatures. The variation of stripe filling with
temperature, also included in this Figure, shows that its
zero temperature value is mostly achieved already at $T\approx t$.
The spin-spin correlation across the stripe ($S_1$) experiences a
sign change, in this case from AF to ferromagnetic, at $T\approx
0.15 t$, after it reaches its maximum AF value which occurs in
turn once spin order in the intervening spin ladders (as shown
along the legs at the maximum distance, $S_2$, and on the rungs,
$S_3$), is well established. In fact, the inter-domain correlation
$S_1$ starts to vary with temperature at a much lower temperature
than the intra-domain ones, $S_2$ and $S_3$. The spin correlation
on stripe rungs ($S_4$) essentially vanishes below the crossover
temperature. These combined features suggest that, at least for the
parameters examined, the formation of anti-phase domain is a result
of a collective interplay between spin domains and not a result of
local correlations on the stripes.

On the other hand, the peak in the charge structure factor at
$(\pi/4,0)$ remains unchanged down from $T\approx 5$ for the same 
parameters of Fig.~\ref{fig4}. At larger temperatures $C({\bf k})$
is virtually constant at nonzero momentum.

\begin{figure}
\begin{center}
\epsfig{file=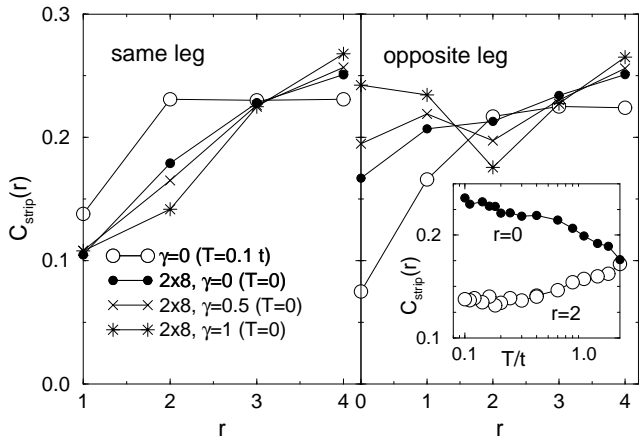,width=6.0cm,angle=-90}
\end{center}
\caption{Hole-hole correlations on the $8\times 8$ cluster,
$x=1/8$, $J=0.7$, $e_s=1.5$, $\gamma=0$. Results obtained
by exact diagonalization on the $2\times 8$ cluster at 
zero temperature are also shown for comparison.
}
\label{fig9}
\end{figure}

Let us examine next hole-hole correlations. In Fig.~\ref{fig9} we
show these correlations, for the same parameters as before, as a
function of distance at the lowest temperature available. We have
also included for comparison the exact correlations obtained for
a $2\times 8$ quarter-filled ladder at $T=0$, with the same
parameters and various values of $\gamma$. Two features are apparent.
First, for $\gamma=0$, correlations on the stripe are quite different
from their zero temperature, isolated ladder, counterparts. In the
inset of this Figure, it is shown that the temperature evolution of 
the stripe correlations is
quite smooth down to the lowest temperature reached. This behavior
suggests that BC stripes do not behave like isolated ladders unless
there is some other crossover at lower temperatures. Second, notice
that in isolated ladders  
pairing occurs near the isotropic limit of the exchange interaction
($\gamma \ge 0.5$) of the $t-J$ model. Unfortunately, it is not
possible to reach low enough temperatures and large
values of $\gamma$ with our QMC algorithm to further investigate these
issues.

\section{Charge transport.}
\label{invkin}

Another important issue a theory of stripes should address is the
behavior of resistivity in the underdoped region. Resistivity
measurements on high-quality La$_{2-x}$Sr$_x$CuO$_4$ (LSCO) and 
YBa$_2$Cu$_3$O$_y$ single crystals\cite{ando2} show a metallic
behavior at moderate temperatures in a wide range of doping. 
These experiments also show a divergent resistivity as
$T\rightarrow 0$ which was
earlier detected in LSCO (Ref.~\onlinecite{ando}) and in
Bi$_2$Sr$_{2-x}$La$_x$CuO$_{6+\delta}$ (Ref.~\onlinecite{ono})
even after suppressing superconductivity with magnetic fields.
The inverse mobility of carriers, also shown in 
Ref.~\onlinecite{ando}, has a similar behavior with temperature
as the resistivity.

A measure of charge mobility is the kinetic energy, which can be
easily computed in QMC and it is affected by relatively small
errors.
In addition, the kinetic energy, due to the use of PBC,
is the most important contribution to the Drude weight (defined
below). Although a straightforward comparison
with the above mentioned experimental results is not possible,
one can get some interesting qualitative insights in the problem.

Further information on charge transport properties can be gained by
looking at the ``paramagnetic" contribution to the Drude
weight.\cite{wagner} To this end, we computed the
current-current correlations in imaginary time defined as:
\begin{eqnarray}
C_{\alpha \alpha}(\tau) = \langle j_{\alpha}(\tau) 
j_{\alpha} \rangle = \sum_{ l,m}  \langle j_{\alpha,l}(\tau)
 j_{\alpha,m} \rangle \label{curcur} \\
=\sum_{ l,m} \frac{1}{\cal Z} Tr \{ j_{\alpha,l}(\tau)
 j_{\alpha,m} e^{-\beta H} \}
\nonumber
\end{eqnarray}
\noindent
where the paramagnetic ${\bf q}=(0,0)$ current operator along 
direction $\alpha$ ($\alpha={\hat x}, {\hat y}$) is:
\begin{eqnarray}
j_{\alpha} = \sum_{ l}  j_{\alpha,l} = i t \sum_{ l,\sigma }
({\tilde c}^{\dagger}_{ l+\alpha,\sigma}
{\tilde c}_{ l,\sigma} - {\tilde c}^{\dagger}_{ l,\sigma}
{\tilde c}_{ l+\alpha,\sigma})
\label{current}
\end{eqnarray}
\noindent
and $j_{\alpha,l}(\tau)=e^{\tau H} j_{\alpha,l} e^{-\tau H}$,
$\tau=l \Delta$, $0\le l < M$.
${\cal Z} = Tr \{ e^{-\beta H} \}$ is the partition 
function. Since $j_{\alpha,l}$ conserves particle number and spin the
calculation of $\langle j_{\alpha,l}(\tau) j_{\alpha,m} \rangle$ 
is relatively simple. Since,
\begin{eqnarray}
\langle j_{\alpha}(\tau) j_{\alpha} \rangle = \frac{1}{\pi}
\int_{-\infty}^{\infty} \frac{\omega {\sigma^\prime_{\alpha}}_{reg}(\omega)} 
{1-e^{-\beta \omega}} e^{-\tau \omega} d\omega
\end{eqnarray}
\noindent
where ${\sigma^\prime_{\alpha}}_{reg}(\omega)$ is the regular part of
the real part of the conductivity along the $\alpha$-direction,
it can be shown that:
\begin{eqnarray}
\chi_{\alpha} = \int_0^{\beta} \langle j_{\alpha}(\tau) j_{\alpha} 
\rangle d\tau=
\frac{2}{\pi} \int_0^{\infty} {\sigma^\prime_{\alpha}}_{reg}(\omega) 
d\omega
\end{eqnarray}

Then, from the standard f-sum rule
one obtains the expression for the Drude weight:\cite{wagner}
\begin{eqnarray}
\frac{2 D_{\alpha}}{\pi} = - \chi_{\alpha} +
\frac{1}{N}\langle K_{\alpha}\rangle
\end{eqnarray}

More elaborated analysis, such as the study
of the dependence of the conductivity with frequency, imply the
analytical continuation of imaginary time current-current
correlations to real frequencies by, for instance, the maximum
entropy method. This procedure is no longer in principle exact
and results would be affected with large error bars.

\begin{figure}
\begin{center}
\epsfig{file=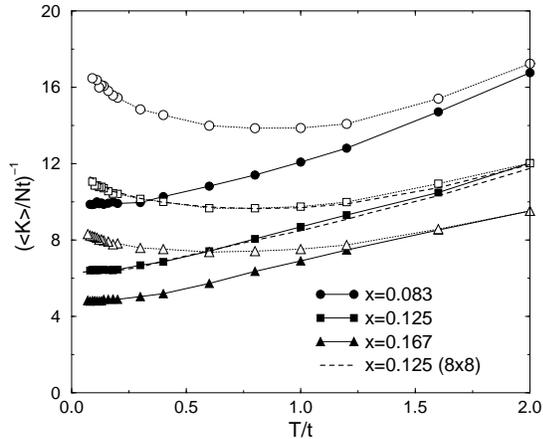,width=6.0cm,angle=-90}
\end{center}
\caption{Inverse kinetic energy per site along (filled symbols) and
perpendicular (open symbols) to the stripes on the $12\times 12$
cluster, $J=0.35$, $e_s=2.0$, and $\gamma=0.0$. Different dopings
are indicated on the plot. Dashed line corresponds to results
obtained on the $8\times 8$ cluster.
}
\label{fig10}
\end{figure}

Let us first examine the inverse kinetic energy per site as a
function of temperature and for various doping fractions in the case
of site centered stripes. In Fig.~\ref{fig10} we show the results of
QMC simulations for the inverse kinetic energy per site along and
perpendicular to the stripes, $\kappa_y=(K_y/N)^{-1}$ and
$\kappa_x=(K_x/N)^{-1}$ resp., on the $12\times 12$ cluster,
$J=0.35$, $e_s=2.0$, for the $t-J_z$ model, and fillings $x=0.083$,
0.125, and 0.167. Both quantities decrease with doping, following the
general trend as the inverse mobilities in Ref.~\onlinecite{ando}.
Most interestingly, it can be seen that although
$\kappa_y$ has a positive slope in the whole temperature range, 
consistent with a metallic behavior of the stripes and in agreement
with the results obtained in the previous section,
$\kappa_x$ changes it sign from positive slope at high temperature
to negative at low temperature indicating charge localization 
in the direction perpendicular to the stripes.

One can determine a temperature scale where $\kappa_x$ and $\kappa_y$
start to separate (within error bars) from each other as $T$ is
lowered. This 
temperature scale coincides roughly with $T_{CO}$ in Fig.~\ref{fig3}.
Another scale, at a lower temperature, can be determined at the
minimum of $\kappa_x$, i.e. at the point where charges start to be
localized in the direction transversal to the stripes.
This scale of temperature,
determined on the $8\times 8$ cluster as a function of $e_s$ has been
plotted in Fig.~\ref{fig3}. From Fig.~\ref{fig10}, this temperature
scale decreases with doping, following the general behavior of the
pseudogap. This temperature scale is located between
$T_{CO}$ and $T_{SO}$ in Fig.~\ref{fig3}.

\begin{figure}
\begin{center}
\epsfig{file=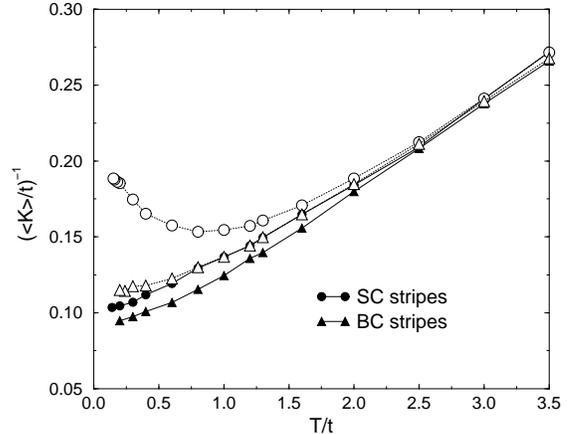,width=6.0cm,angle=-90}
\end{center}
\caption{Inverse kinetic energy per site along (filled symbols) and
perpendicular (open symbols) to the stripes on the $8\times 8$
cluster, $J=0.7$, $e_s=2.0$, $\gamma=0$ for site centered and bond
centered stripes.
}
\label{fig11}
\end{figure}

As expected, the effect of the anisotropy in the exchange term of the
Hamiltonian is very small for $\kappa_x$ and $\kappa_y$. For 
$\gamma=0$ and 0.5, results practically coincide within error bars.
On the other hand, it is instructive to compare the behavior for
SC and BC stripes. In Fig.~\ref{fig11}, we show the inverse 
kinetic energies obtained for the $8\times 8$ cluster, $J=0.7$,
$e_s=2.0$, $\gamma=0$, for both types of stripes. These results show
a larger mobility both along and transversal to the stripes and
an absence of localization in the case of BC stripes down to the
lowest temperature reached ($T=0.2 t$).

\begin{figure}
\begin{center}
\epsfig{file=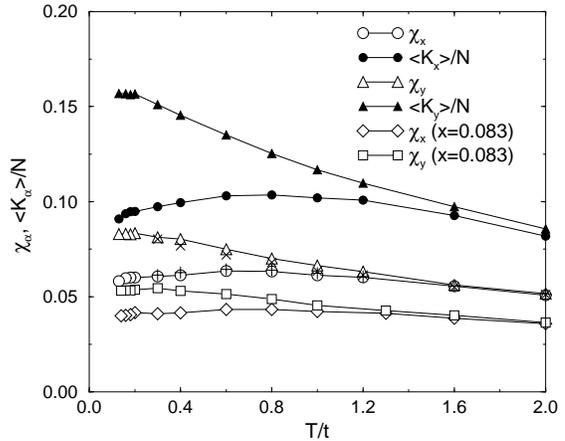,width=6.0cm,angle=-90}
\end{center}
\caption{Current susceptibility and kinetic energy per site along 
and perpendicular to the stripes on the $8\times 8$ cluster,
$x=0.125$, $J=0.35$, $e_s=2.0$, $\gamma=0$ for site centered
stripes. $\chi_x$ and $\chi_y$ on the $12 \times 12$ cluster and for
the same doping are indicated with ``$+$" and ``$\times$"
respectively. $\chi_x$ and $\chi_y$ on the $12 \times 12$ cluster
for $x=0.083$ are also shown.
}
\label{fig12}
\end{figure}

The behavior of the current susceptibility $\chi_{\alpha}$ is very
similar to the kinetic energy per site along the same direction.
Typical results are shown in Fig.~\ref{fig12} obtained on the
$8\times 8$ cluster for $x=0.125$, $J=0.35$, $e_s=2.0$, $\gamma=0$.

\noindent
Two differences between those two quantities are
apparent. First,
the separation between the
kinetic energies parallel and perpendicular to the stripes are
larger than for the corresponding $\chi_y$ and $\chi_x$.
Second, both temperature scales discussed above are larger for
the kinetic energy per site than for the current susceptibility.
Both features, which we have seen in all cases examined, suggest
that the diamagnetic response is more sensitive to charge
inhomogeneity than the paramagnetic one. Notice also that 
finite size effects for these current susceptibilities are very
small.

\section{Conclusions}

We have examined various properties of a two-dimensional $t-J$
model in which charge inhomogeneity is stabilized by an on-site
potential by using two different numerical techniques.
Diagonalization in a restricted Hilbert space and finite
temperature Quantum Monte Carlo techniques allow us to work with
relatively large clusters with fully periodic boundary conditions.
Extrapolation to the full Hilbert space and extrapolation to zero
temperature give quite consistent results.

In the first place, we obtained that the linear filling in BC stripes
is larger than in SC stripes. In this latter case, definitely,
a stabilizing mechanism is needed if this model has to reproduce
experimental results.

Then, we showed that the effect of a moderate XY term in the
Heisenberg interaction does not affect the main conclusions obtained
in the Ising limit in Ref.~\onlinecite{riera} for SC stripes.
First, the anti-phase domain ordering, which generates the IC
peaks measured in neutron scattering experiments occurs at a
much lower temperature than the formation
of charge inhomogeneities and charge localization. This magnetic
ordering is a consequence of a collective interplay between the
spin domains and are not driven by local correlations located on
the stripes. Second, hole-hole correlations indicate
a metallic behavior of the stripes with no signs of hole attraction.
The study of the doping dependence in the range $0.083 \le x\le
0.167$ suggest that these features are characteristic of the
whole underdoped region.

In addition, we showed that some of the main conclusions for SC
stripes apply to BC stripes as well, although in this case our study
was mostly limited to the Ising limit of the Hamiltonian.
Again, as the temperature decreases
the sequence of events are: first, the formation of charge
inhomogeneities, then the establishment of AF order in the
intervening spin domains and finally the $\pi$-shifted magnetic
ordering of the whole system. The hole-hole correlations on the
stripe show again a metallic behavior and, as a difference with
SC stripes, which are very similar to isolated chains, they indicate
that, at finite temperatures, stripes do not behave quite like 
isolated ladders, unless another crossover occurs at lower
temperatures than the ones we can reach.

Finally, the study of charge transport properties provides further
support to the physical validity of the model defined by
Eqs.~(\ref{ham_anis})-(\ref{stripepot}). In particular, in the
case of site centered stripes, we have shown distinctive
behavior of the inverse kinetic energy, as a measure of charge
mobility, in the directions parallel and perpendicular to the
stripes, in this latter case showing a characteristic feature
of localization. A similar behavior is observed in the
paramagnetic part of the Drude weight computed through the
current-current correlations. In the case of bond centered stripes,
on the other hand, the increase of the inverse kinetic energy
as $T \rightarrow 0$ has not been observed down to the lowest
attainable temperature. Although these results were obtained near
the Ising limit and for well-defined charge inhomogeneities,
we believe they could help to experimentally decide between
SC and BC stripes in underdoped cuprates.

\acknowledgements
I wish to acknowledge many interesting discussions
with A. Dobry and A. Greco. I also thank D. Poilblanc
for running some of the codes on the NEC supercomputer
at IDRIS, Orsay (France).

\end{document}